\documentclass[prd,nofootinbib,tightenlines,showkeys,showpacs]{revtex4}
\usepackage{amsfonts}
\usepackage{amssymb}
\usepackage{amsmath}
\usepackage{indentfirst}
\usepackage[latin1]{inputenc}

\begin{document}
\title{{\small\hfill IISc-CHEP/01/09}\\
{\small\hfill IMSC-2009/01/01}\\
{\small\hfill SU-4252-883}\\
Spontaneous Symmetry Breaking In Twisted Noncommutative Quantum
  Theories}

\author{A. P. Balachandran\footnote{bal@phy.syr.edu}}
\affiliation{{C\'atedra de Excelencia,}
Departamento de Matem\'atica, Universidad Carlos III de Madrid,
28911, Legan\'es, Madrid, Spain \\
and \\
Physics Department, Syracuse University
Syracuse, NY, 13244-1130, USA }
\author{T. R. Govindarajan\footnote{trg@imsc.res.in}}
\affiliation{The Institute of Mathematical Sciences
C. I. T. Campus Taramani, Chennai 600 113, India
}
\author{Sachindeo Vaidya\footnote{vaidya@cts.iisc.ernet.in}}
\affiliation{Centre for High Energy Physics,
Indian Institute of Science, Bangalore 560 012, India
}

\pacs{ 11.10.Nx, 11.30.Cp}

\keywords{Gauge fields, Non-Commutative Geometry, twisted Poincar\'e symmetry}

\begin{abstract}
We analyse  aspects of symmetry breaking for Moyal spacetimes
within a quantisation scheme which preserves the twisted Poincar\'e symmetry. 
Towards this purpose, we develop the LSZ approach for Moyal spacetimes. The 
latter gives a formula for scattering amplitudes on these 
spacetimes which can be 
obtained from the corresponding ones on the commutative spacetime. 
This formula applies in the presence of spontaneous breakdown of symmetries as well. 
We also derive Goldstone's theorem on Moyal spacetime. The formalism developed
here can be directly applied to the twisted standard model.
\end{abstract}

\maketitle

\section{Introduction}

Spontaneous breaking of a continuous symmetry involves a subtle interplay
between an infinite number of degrees of freedom, local and spacetime
symmetries, dimension of spacetime, and the notion of (non-)locality
of interaction. Naturally one would suspect that the phenomenon of
spontaneous symmetry breaking (SSB) leads to different physics in the
context of quantum field theories on the Groenewold-Moyal (GM) plane,
when the idea of locality is altered, albeit in a very precise sense:
pointwise multiplication of two functions is replaced by {\it
star}-multiplication, which is no longer commutative, and is in
addition non-local. New phases 
and soliton solutions appear making the dynamics 
richer\cite{gubser,stripe,trg1,trg2}.

Writing quantum field theories on such spaces requires some care, if
one wishes to discuss questions related to Poincar\'e invariance. To
give up this spacetime symmetry almost entirely (which is what
conventional quantization does) seems too heavy a price, as it affects
the notion of identity of particles (``of two identical particles in
one frame should describe two identical particles in {\it all}
reference frames''), and leads to unacceptable coupling between UV-
and IR- degrees of freedom as well\cite{minwalla}. The program of twisted
quantization initiated in \cite{bmpv,replyto} on the the other hand,
avoids these pitfalls: Poincar\'e invariance can be maintained, a
generalized notion of identical particles can be defined, and UV and
IR degrees of freedom decouple nicely \cite{bpq}, thus rekindling the
hope that phenomenologically interesting models can be
constructed. Indeed one can construct quantum gauge theories with
arbitrary gauge groups consistently \cite{bpqv1}.

In this paper, we address the issue of SSB and Higgs phenomenon in
twisted quantum theories, and demonstrate signatures of
noncommutativity. Our general formulation applies to an
arbitrary group $G$ breaking to a subgroup $H$. The extension
to the (noncommutative) Standard Model and beyond-Standard Model
physics is conceptually straightforward, and will be discussed as
well. Such physics merits a more elaborate investigation which we
reserve for later work.

This paper is organised as follows. In Section 2, we will review
twisted quantization on noncommutative spaces and gauge theories based
on this formalism. Section 3 will elaborate the LSZ formalism for
twisted quantisation and discuss in detail the Gell-Mann-Low formula and
its modifications on the GM plane.  In Section 4, we will summarise
our rules for twisted quantum field theories followed by application
of the same to spontaneously broken theories and Higgs mechanism on the 
GM plane in Sec.5.  Our conclusions and future outlook are in Section
6.

\section{Twisted Quantization}

The algebra of functions ${\cal A}_\theta({\mathbb R}^N)$ on the GM
plane consists of smooth functions on ${\mathbb R}^N$, with the
multiplication map
\begin{eqnarray}
m_\theta: {\cal A}_\theta ({\mathbb R}^N) \otimes {\cal A}_\theta
({\mathbb R}^N) &\rightarrow& {\cal A}_\theta ({\mathbb R}^N)\,,
\nonumber \\
\alpha \otimes \beta &\rightarrow& \alpha \;e^{\frac{i}{2}
  \overleftarrow{\partial}_\mu \theta^{\mu \nu}
  \overrightarrow{\partial}_\nu} \ \beta := \alpha \ast \beta
\label{starmult}
\end{eqnarray}
where $\theta^{\mu \nu}$ is a constant antisymmetric tensor.

Let
\begin{equation}
F_\theta = e^{\frac{i}{2} \partial_\mu \otimes \theta^{\mu \nu}
  \partial_\nu} = {\rm ``Twist \; element"}.
\label{twistelt}
\end{equation}
Then
\begin{equation}
m_\theta (\alpha \otimes \beta) = m_0 [F_\theta \alpha \otimes
\beta] \label{starmult1}
\end{equation}
where $m_0$ is the point-wise multiplication map, also defined by
(\ref{starmult}).

The usual action of the Lorentz group ${\cal L}$ is not compatible
with $\ast$-multiplication: transforming $\alpha$ and $\beta$
separately by an arbitrary group element $\Lambda \in {\cal L}$ and
then $\ast$-multiplying them is not the same as transforming their
$\ast$-product. In other words, the usual coproduct $\Delta_0(\Lambda)
= \Lambda \otimes \Lambda$ on the group algebra ${\mathbb C}\cal{L}$ of $\cal{L}$
is not compatible with $m_\theta$. But a new
coproduct $\Delta_\theta$ obtained using the twist is compatible,
where
\begin{equation}
\Delta_\theta(\Lambda)\,=\,F_\theta^{-1}\,\Delta_0 (\Lambda) F_\theta.
\label{twistedcoprod}
\end{equation}
Here $\partial_\mu \otimes \theta^{\mu \nu}\partial_\nu$ in $F_\theta$
is to be replaced by $~-~ P_\mu \otimes \theta^{\mu \nu} P_\nu$ where
$P_\mu$ are translation generators: we are dealing with
${\cal{P}}_\theta \otimes {\cal{P}}_\theta$ where ${\cal{P}}_\theta$
is a Hopf algebra associated with the Poincar\'e group algebra ${\mathbb C}{\cal{P}}$ with
the coproduct (\ref{twistedcoprod}).

This twisted coproduct does not preserve standard
(anti-)symmetrization, because it does not commute with the usual flip
operator $\tau_0$ defined by $\tau_0:(\alpha \otimes \beta) ~=~
(\beta\otimes\alpha)$:

\begin{equation}
\Delta_\theta(\Lambda) \tau_0 \neq \tau_0 \Delta_\theta (\Lambda).
\end{equation}

On the other hand, it does preserve twisted (anti-)symmetrization,
defined using a new flip operator $\tau_\theta = F_\theta^{-1}\,\tau_0
(\Lambda) F_\theta$:
\begin{equation}
\Delta_\theta(\Lambda) \tau_\theta = \tau_\theta \Delta_\theta (\Lambda).
\end{equation}

Thus in noncommutative quantum theory, the usual fermions/bosons do
not make sense, but twisted ones do. They are obtained from the
projectors $S_\theta, A_\theta$:
\begin{equation}
S_\theta~=~\frac{{\bf 1}~+~\tau_\theta}{2}, \qquad A_\theta~=~ 
\frac{{\bf 1}~-~\tau_\theta}{2}.
\end{equation}

\subsection{Quantum Fields}

A quantum field $\chi$ on evaluation at a spacetime point (or more generally
on pairing with a test function) gives an operator acting on a Hilbert
space. A field at $x_1$ acting on the vacuum gives a one-particle
state centered at $x_1$. When we write $\chi(x_1)\,\chi(x_2)$ we mean
$(\chi\otimes\chi)(x_1,x_2)$. Acting on the vacuum we generate a
two-particle state, where one particle is centered at $x_1$ and the
other at $x_2$.

If $a_p$ is the annihilation operator of the free second-quantized field
$\phi_\theta$ on ${\cal A}_\theta({\mathbb R}^N)$, we want, as in standard quantum field theory,
\begin{eqnarray} 
\langle 0 |\phi_\theta(x) a^\dagger_p |0\rangle &=& e_p(x), \\
\frac{1}{2}\langle 0 |\phi_\theta(x_1) \phi_\theta(x_2) a^\dagger_q
a^\dagger_p |0\rangle &=& \left(\frac{{\bf 1} \pm
  \tau_\theta}{2}\right)(e_p \otimes e_q)(x_1,x_2) \nonumber \\
&\equiv& (e_p \otimes_{S_\theta,A_\theta} e_q)(x_1,x_2) 
\label{tbasis} 
\end{eqnarray} 
where $e_p(x) = e^{-i p \cdot x}$.

This compatibility between twisted statistics and Poincar\'e
invariance has profound consequences for commutation relations. For
example when the states are labeled by momenta, we have, from
exchanging $p$ and $q$ in (\ref{tbasis}),
\begin{equation}
|p, q\rangle_{S_\theta,A_\theta} =\ \pm\,e^{ i \theta^{\mu\nu}p_\mu
 q_\nu}\,|q,p \rangle_{S_\theta,A_\theta}. 
\end{equation}
This is the origin of the commutation relations
\begin{eqnarray}
a_p^\dagger\,a_q^\dagger\,&=& \pm e^{ i \theta^{\mu\nu}p_\mu
  q_\nu}\,a_q^\dagger\,a_p^\dagger \, , \\ 
a_p a_q &=&  \pm e^{ i \theta^{\mu \nu} p_\mu q_\nu} a_q a_p \, .
\end{eqnarray}

\subsection{Gauge Theories}

The algebra ${\cal A}_\theta({\mathbb R}^N)$, regarded as a vector
space, is a module for ${\cal A}_0({\mathbb R}^N)$. This observation
is of central importance to us, as it allows us to
write gauge theories based on arbitrary gauge groups (as opposed to
just $U(N)$). We can show this as follows.

For any $\alpha \in {\cal A}_\theta({\mathbb R}^N)$, we can define two
representations $\hat{\alpha}^{L,R}$ acting on ${\cal
A}_\theta({\mathbb R}^N)$:
\begin{equation}
\hat{\alpha}^L \xi = \alpha * \xi, \quad \hat{\alpha}^R \xi = \xi
* \alpha \quad {\rm for} \quad \xi \in {\cal A}_\theta({\mathbb
R}^N) \ ,
\end{equation}
where $*$ is the GM product defined by Eq.(\ref{starmult}) (or,
equivalently, by Eq.(\ref{starmult1})).  The maps $ \alpha \rightarrow
\hat{\alpha}^{L,R}$ have the properties
\begin{eqnarray}
\hat{\alpha}^L \hat{\beta}^L &=& (\hat{\alpha}\hat{\beta})^L, \\
\hat{\alpha}^R \hat{\beta}^R &=& (\hat{\beta}\hat{\alpha})^R, \label{right}\\
{[}\hat{\alpha}^L, \hat{\beta}^R] &=& 0. \label{LRcommute}
\end{eqnarray}
The reversal of $\hat{\alpha}, \hat{\beta}$ on the right-hand side of
(\ref{right}) means that for position operators,
\begin{equation}
{[}\hat{x}^{\mu L}, \hat{x}^{\nu L}] = i \theta^{\mu \nu} =
-[\hat{x}^{\mu R}, \hat{x}^{\nu R}].
\end{equation}
Hence in view of (\ref{LRcommute}),
\begin{equation}
\hat{x}^{\mu c} = \frac{1}{2} \left( \hat{x}^{\mu L} + \hat{x}^{\mu R}
\right)
\end{equation}
generates a representation of the commutative algebra ${\cal
  A}_0({\mathbb R}^N)$:
\begin{equation}
{[}\hat{x}^{\mu c}, \hat{x}^{\nu c}] = 0.
\end{equation}

We regard elements of the gauge group ${\cal G}$ as maps from ${\cal
A}_0({\mathbb R}^N)$ to the Lie group $G$.
\begin{equation}
g:~\hat{x^c}\longrightarrow g(\hat{x^c})~\in G
\end{equation}
In cases of interest, $G$ is a compact connected Lie group . For
convenience, we also think of $G$ concretely in terms of the defining
finite-dimensional faithful representation by linear operators.

Fields which transform non-trivially under ${\cal G}$ are modules over
${\cal A}_\theta ({\mathbb R}^N)$. If a $d$-dimensional representation
of $G$ is involved, they can be elements of ${\cal A}_\theta ({\mathbb
R}^N) \otimes {\mathbb C}^d$. Compatibility of gauge transformations
(on these modules) with the $\ast$-product requires us to twist the
coproduct on the gauge group too. The new coproduct is
\begin{equation}
\Delta_\theta (g(\hat{x}^c)) = F_\theta^{-1} [g(\hat{x}^c) \otimes
  g(\hat{x}^c)] F_\theta \;.
\label{twistedgauge}
\end{equation}

This deformation of the coproduct for gauge transformations is
neccessary if we want to be able to construct gauge scalars, and other
composite operators (see \cite{bpqv1} for details).

Finally, we need to understand how to define covariant derivative
$D_\mu$. To this end, consider for simplicity a free charged scalar
field $\phi_\theta(x)$,
\begin{equation} 
\phi_\theta(x) = \int d\mu(p) (a_p e^{-i p \cdot x} + b^\dagger_p
  e^{i p \cdot x}),\quad {\rm where} \quad d \mu(p) \equiv \frac{d^3
  p}{2p_0}, \quad p_0=\sqrt{\vec{p}^2 + m^2}.
\end{equation} 
We require that the field $\phi_\theta$ obeys twisted statistics in
Fock space:
\begin{eqnarray}  
a_p a_q &=& e^{i p \wedge q} a_q a_p , \quad {\rm where} \quad p
\wedge q \equiv p_\mu \theta^{\mu\nu} q_\nu,   \nonumber \\
a_p a^\dagger _q &=& e^{-i p \wedge q} a^\dagger _q a_p + 2p_0
\delta^{(3)}(p-q)
\end{eqnarray} 
and similarly for $b(p), b^\dagger(p)$. The twisted operators $a(p),
a^\dagger(p),b(p),b^\dagger(p)$ can be realized in terms of untwisted Fock space
operators $c(p),d(p)$ and their adjoints through the well-known ``dressing transformations"
\cite{grossefaddeev}
\begin{eqnarray} 
a_p &=& c_p e^{-\frac{i}{2} p \wedge P}, \qquad \qquad a^\dagger _p= c^\dagger _p
e^{\frac{i}{2} p \wedge P}, \nonumber \\
b_p &=& d_p e^{-\frac{i}{2} p \wedge P}, \qquad \qquad b^\dagger _p= d^\dagger _p
e^{\frac{i}{2} p \wedge P}, {\rm where} \\
P_\mu &=& \int d\mu(p) p_\mu \left(a^\dagger_p a_p + b^\dagger_p
b_p \right) = \int d\mu(p) p_\mu \left(c^\dagger_p c_p + d^\dagger_pd_p \right)\nonumber \\
&=& {\rm total \,
  momentum\, operator}.
\end{eqnarray} 
Then $\phi_\theta(x)$ may be written in terms of the ordinary or commutative
fields $\phi_0$ as
\begin{equation} 
\phi_\theta(x) = \phi_0(x) e^{\frac{1}{2}\overleftarrow{\partial} \wedge P},
\quad \overleftarrow{\partial} \wedge P \equiv
\overleftarrow{\partial}_\mu \theta^{\mu\nu} P_\nu
\label{twistphi}
\end{equation} 

As ${\cal A}_\theta({\mathbb R}^N)$ is a module for ${\cal
  A}_0({\mathbb R}^N)$, we require that the covariant derivative
  respects this property. In addition, we also require that in quantum
  theory, the covariant derivative preserves statistics, and also
  Poincar\'e and gauge symmetries. The only possibility that satisfies
  these requirements is
\begin{equation} 
D_\mu \phi_\theta = ((D_\mu)_0 \phi_0)e^{\frac{1}{2} \overleftarrow{\partial}
  \wedge P} .
\label{twistD}
\end{equation} 
where $(D_\mu)_0 = \partial_\mu + (A_\mu)_0$ and $(A_\mu)_0$ is the
commutative gauge field, depending on $\hat{x}^c$ only.

The commutator of two covariant derivatives gives us the curvature:
\begin{equation} 
[D_\mu , D_\nu]\phi = ([(D_\mu)_0, (D_\nu)_0])e^{\frac{1}{2}
  \overleftarrow{\partial} \wedge P} = [(F_{\mu\nu})_0 \phi_0]e^{\frac{1}{2}
  \overleftarrow{\partial} \wedge P}.
\end{equation} 
 
The field strength $(F_{\mu\nu})_0$ transforms correctly
(i.e. covariantly) under gauge transformations, so we can use it to
construct the Hamiltonian of the quantum gauge theory. Pure gauge
theories on the GM plane are thus identical to their commutative
counterparts.

Below we will outline an approach to scattering theory of twisted
fields based on the LSZ formalism. In that approach, (\ref{twistphi})
and (\ref{twistD}) are true in the fully interacting case as
well. Thus these equations are valid with $P_\mu$ being the {\it
total} four momentum of all fields including interactions.

We will discuss spontaneous symmetry breakdown in the presence of
twists later.  Here we remark only the following. If the connected,
compact Lie group $G$ is the gauged group and it is spontaneously
broken to the gauge theory of the subgroup $H$, then the vector fields
acquiring mass are to be twisted.  Only the gauge fields of $H$ escape
the twist.

The LSZ approach to the scattering theory of such interactions appears
to be more streamlined and non-perturbative as compared to our earlier
treatments\cite{bpq}. There we used the interaction representation. In
this representation, the free Hamiltonian is not twisted whereas the
interaction Hamiltonian is
\begin{equation} 
H_\theta^I = \int d^3 x [{\cal H}_\theta^{M-G} + {\cal H}_\theta^{G}]
\end{equation} 
Here ${\cal H}_\theta^{M-G}$ and ${\cal H}_\theta^{G}$ correspond to
the interaction Hamiltonian densities with matter and gauge fields,
and with gauge fields alone respectively. The $\theta$- dependence of
the interaction representation S-matrix disappears when
${\cal{H}}^G_\theta~=~0$, but that is not the case when both
${\cal{H}}_\theta ^M$ and ${\cal{H}}_\theta^G$ are present.
Scattering processes that involve cross terms between ${\cal
H}_\theta^{M-G}$ and ${\cal H}_\theta^{G}$, like the $qg \rightarrow
qg$ scattering in QCD, show effects of noncommutativity.

As we will later point out, for reasons not well understood, for
$\theta ^{\mu\nu}\neq 0$, the LSZ S-matrix differs from the
interaction representation S-matrix and leads to different cross
sections.

\section{The LSZ Formalism for Twisted Quantum Theories}
In the next two subsections below, we review scattering theory,
including the LSZ formalism for standard (untwisted fields). We then
generalise the discussion to the twisted case.
\subsection{Formal scattering theory}
In standard scattering theory, the Hamiltonian $H$ is split into a
``free'' Hamiltonian $H_0$ and an ``interaction'' piece $H_I$,
\begin{equation} 
H = H_0 + H_I,
\end{equation} 
and $H_0$ is used to define the states in the infinite past and
future. Then the states at $t=0$ which in the infinite past (future)
become states evolving by $H_0$ are the in (out) states:
\begin{eqnarray} 
e^{-i H T_-}|\psi; {\rm in}\rangle &\stackrel{T_- \rightarrow -\infty}
\longrightarrow& e^{-i H_0 T_-} |\psi; {\rm F}\rangle, \quad {\rm F}
\equiv {\rm free} \\
e^{-i H T_+}|\psi; {\rm out}\rangle &\stackrel{T_+ \rightarrow +\infty}
\longrightarrow& e^{-i H_0 T_+} |\psi; {\rm F}\rangle.
\end{eqnarray} 

Hence
\begin{eqnarray} 
|\psi; {\rm in}\rangle &=& \Omega_+ |\psi;{\rm F}\rangle, \\
|\psi; {\rm out}\rangle &=& \Omega_- |\psi;{\rm F}\rangle, \\
\Omega_\pm &\equiv& e^{i H T_\mp} e^{-i H_0 T_\mp}, \quad {\rm as} \quad
T_\mp \rightarrow \mp \infty, \\
&=& {\rm M\phi ller}~{\rm operators}.\nonumber
\end{eqnarray} 

We see that 
\begin{eqnarray} 
e^{i H \tau} \Omega_\pm &=& \lim_{T_\mp \rightarrow \mp \infty} e^{i H
  T_\mp } e^{-i H_0 (T_\mp~-~\tau)} , \\
&=& \Omega_\pm e^{i H_0 \tau}
\end{eqnarray} 
and
\begin{equation} 
|\psi; {\rm out} \rangle = \Omega_- \Omega_+^\dagger |\psi; {\rm in}\rangle
\end{equation} 
If the incoming state is $|k_1, k_2, \cdots k_N; {\rm F}\rangle$, it
follows that
\begin{equation} 
|k_1, k_2, \cdots k_N; {\rm in}\rangle = \Omega_+ |k_1, k_2, \cdots
 k_N; {\rm F}\rangle 
\end{equation} 
has eigenvalue $\sum k_{i0}$ for the {\it total} Hamiltonian $H$. A
similar result is true for
\begin{equation} 
|k_1, k_2, \cdots k_N; {\rm out}\rangle = \Omega_- |k_1, k_2, \cdots
 k_N; {\rm F}\rangle .
\end{equation}
We note that the scattering amplitude is 
\begin{equation} 
\langle \xi; {\rm out} | \psi; {\rm in}\rangle = \langle \xi; {\rm
  in}|\Omega_+ \Omega_-^\dagger |\psi; {\rm in}\rangle.
\end{equation} 
In other words, the LSZ $S$-matrix is 
\begin{equation} 
S = \Omega_+ \Omega_-^\dagger, \quad |\psi; {\rm out} \rangle =
S^\dagger |\psi; {\rm in}\rangle.
\end{equation} 
Between the ``free'' states, the $S$-operator is different:
\begin{equation} 
\langle \xi; {\rm out} | \psi; {\rm in}\rangle = \langle \xi; {\rm
  F}|\Omega_-^\dagger  \Omega_+ |\psi; {\rm F}\rangle.
\end{equation} 

The LSZ formalism works exclusively with the in- and out-states, as
Haag's theorem shows that $\Omega_\pm$ do not exist in quantum field
theories.  The creation-annihilation operators $c_k^{\rm in (out)
\dagger}, c_k^{\rm in (out)}$ are introduced to create states
$|k_1,k_2,\cdots k_N; {\rm in(out)}\rangle$ from the vacuum. The in-
and out- fields $\phi_{\rm in(out)}$ are then defined using
superposition. They look like free fields, but are not, since for the
{\em total} four-momentum $P_\mu$, we have
\begin{equation} 
P_\mu |k_1,k_2,\cdots k_N; {\rm in(out)}\rangle = (\sum_i k_{i \mu})
|k_1,k_2,\cdots k_N; {\rm in(out)}\rangle .
\end{equation} 

It is also assumed that 

(a) The vacuum and single particle states are unique up to a
phase. Then after a phase choice, there is only one vacuum
$|0\rangle$, $\langle 0 | 0\rangle =1$, and
\begin{equation} 
S|0\rangle = |0\rangle.
\end{equation} 

(b) There exists an interpolating field $\phi$ in the Heisenberg
representation such that matrix elements of $\phi(x,t)$ between in-
and out- states go to those of $\phi_{\rm in, out}(x,t)$ in the
infinite past and future,
\begin{equation} 
\phi(x,t)~ -~ \phi_{\rm in, out}(x,t) \rightarrow 0 \quad {\rm as} \quad t
\rightarrow \mp \infty 
\end{equation} 
in weak topology. (We treat the case of just one scalar field for simplicity.)

Then LSZ formalism shows that 
\begin{equation} 
\langle k'_1,k'_2,\cdots k'_N; {\rm out}|k_1,k_2,\cdots k_M; {\rm
  in}\rangle = \int {\cal{I}}~~G_{N+M}(x_1',\cdots,x_N';x_1\cdots x_M)
\end{equation} 
where 
\begin{equation}
{\cal{I}}~=~\prod d^4x'_i\prod d^4x_j~e^{i(k_i'\cdot
x_i'~-~k_j\cdot x_j)} i(\partial_i'^2+m^2)\cdot i(\partial_j^2+m^2)
\label{I}
\end{equation}
and 
\begin{equation}
G_{N+M} (x_1',\cdots,x_N';x_1\cdots x_M)\equiv \langle
0|T(\phi(x_1')\cdots \phi(x_N')\phi(x_1)\cdots \phi(x_M))|0\rangle
\end{equation}

It is now convenient to regard all the momenta as ingoing, relabel them as $q_1,q_2,\cdots q_{N+M}$
and write 
\begin{equation}
\langle -q_1,q_2, \cdots -q_N; {\rm out} \mid q_{N+1},\cdots q_{N+M}; {\rm in}\rangle~=~\int {\cal{I}}~G_{N+M},
\end{equation}
where 
\begin{equation}
{\cal{I}}~=~\prod^{N+M}_{i=1} d^4x_i~e^{-iq_i\cdot x_i}~i(\partial_i^2+m^2)
\label{II}
\end{equation}
and
\begin{equation}
G_{N+M} (x_1,\cdots,x_{N+M})~=~\langle 0\mid T(\phi(x_1)\cdots \phi(x_{N+M})\mid 0 \rangle.
\label{LSZformula}
\end{equation}

We will later see the differences in the scattering amplitude on the
GM plane through an analysis of the Gell-Mann-Low formula.
\subsection{The Gell-Mann-Low Formula}
The Heisenberg fields $\phi$ and the free fields $\phi_F$ at time $t~=~0$
fulfill the {\it same} canonical algebra if the interaction has no
time derivatives. We assume that to be the case.

Then in perturbation theory, we choose the same representation of the
canonical algebra at time 0, namely that of the free field $\phi_F$:
\begin{equation} 
\phi(\cdot, 0) = \phi_F(\cdot,0) .
\end{equation} 
This implies that
\begin{eqnarray} 
\phi_F(\cdot, t) &=& e^{iH_0 t}\phi_F(\cdot,0)e^{-i H_0 t}, \\
\phi(\cdot,t) &=& e^{i H t}\phi(\cdot,0)e^{-i H t} = e^{i H
  t}\phi_F(\cdot,0)e^{-i H t} 
\end{eqnarray} 
or
\begin{equation} 
\phi(\cdot,t) = (e^{i H t}e^{-i H_0 t} \phi_F(\cdot,t)(e^{i H_0 t}
e^{-i H t})
\end{equation} 
Let us define
\begin{equation} 
U(t_1,t_2) = e^{i t_1 H_0}e^{-i(t_1-t_2)H}e^{-it_2 H_0}
\end{equation} 
Then
\begin{eqnarray} 
U(t,t) &=& 1, \\
i \frac{\partial U}{\partial t_1}(t_1,t_2) &=& H_I(t_1) U(t_1,t_2),
\end{eqnarray} 
where
\begin{equation} 
H_I(t) = e^{it_1 H_0} H_I(0) e^{-i t_1 H_0}, \quad H_I(0) = H_I
\end{equation} 
is the interaction representation Hamiltonian. 

Thus
\begin{equation} 
U(t_1,t_2) = T \exp \left( -i \int_{t_1}^{t_2} dt H_I (t) \right)
\end{equation} 
and 
\begin{equation} 
\phi(\cdot,t) = U(0,t) \phi_F(\cdot,t) U(t,0)
\label{heisenberg}
\end{equation} 

Gell-Mann and Low show that 
\begin{equation} 
G_{N+M}(x_1,\cdots x_{N+M})= \frac{\langle 0,{\rm F}|T \left(\phi_F(x_1) \cdots
  \phi_F(x_{N+M}) e^{i\int d^4 x {\cal L}_I (x)}\right) |0,{\rm
  F}\rangle}{\langle 0,{\rm F}| e^{i\int d^4 x {\cal L}_I (x)}|0,{\rm
  F}\rangle}
\end{equation} 
where $|0,{\rm F}\rangle$ is the vacuum of the free Hamiltonian $H_0$:
\begin{equation} 
H_0 |0, {\rm F}\rangle = 0 .
\end{equation} 

The proof is standard and will be omitted here.

\subsection{The twisted quantum fields}
Let us look at the case when $\theta^{\mu\nu} \neq 0$, first focussing
on the situation with no gauge fields. Gauge fields will be treated later. 

Our assumption is that the noncommutative field theory comes from the
commutative one by the replacement
\begin{equation} 
\phi_\theta = \phi_0 e^{\frac{1}{2} \overleftarrow{\partial} \wedge
  P}. 
\label{twistfield}
\end{equation} 
for {\it matter} fields, whereas gauge fields are {\it not} twisted (See also 
(\ref{twistD})).
As $t\rightarrow ~\pm\infty$, the Heisenberg field $\phi_0$ for
$\theta_{\mu\nu}~=~0$ becomes the corresponding in- and out- fields
$\phi_0^{in,out}$.  As for $P_\mu$, they are not affected by these
limits, being constants of motion. Hence formally, we find, for the
in- and out- fields $\phi^{in,out}_\theta$ of $\phi_\theta$, the
results
\begin{equation} 
\phi_\theta \rightarrow \phi_\theta^{\rm in,out}~:~= \phi_{\rm in, out}
e^{\frac{1}{2}\overleftarrow{\partial} \wedge P}
\quad {\rm as} \quad t \rightarrow \mp \infty.
\label{phiin}
\end{equation}
For the twisted in and out annihilation and creation operators
$a_k^{in,out},a_k^{\dagger in,out}$ we thus find 
\begin{equation}
a_k^{in,out}~=~c_k^{in,out}e^{-\frac{i}{2}k_\mu\theta^{\mu\nu}P_\nu},\qquad
a_k^{\dagger in,out}~=~c_k^{\dagger in,out}e^{\frac{i}{2}k_\mu\theta^{\mu\nu}P_\nu}.\label{inout}
\end{equation}

There is a further convention we want to explain. For consistency with our notation 
for the coproduct on the Poincar\'e group\cite{07081379}, we define 
\begin{equation}
a_k^{\dagger in}|k_r,k_{r-1},\cdots ,k_1\rangle_{\rm in}~=~ |k_r,k_{r-1},\cdots ,k_1,k\rangle_{\rm in}
\end{equation}
and similarly for the action of $a_k^{\dagger out}$. Thus for example
\begin{equation}
|k_r,k_{r-1},\cdots ,k_1\rangle_{\rm in}~=~a_{k_1}^{\dagger in}a_{k_2}^{\dagger in}\cdots 
a_{k_r}^{\dagger in}|0\rangle_{\rm in} .
\end{equation}
and 
\begin{equation}
\langle -q_N,-q_{N-1},\cdots -q_1;out|q_{N+M},q_{N+M-1},\cdots ,q_{N+1}\rangle
~=~\int {\cal{I}}~G_{N+M}^\theta(x_1,x_2,\cdots , x_{N+M})
\label{scatter}
\end{equation}
$G_{N+M}~\equiv G_{N+M}^0$ has got changed to $G_{N+M}^\theta $ for $\theta_{\mu\nu}\neq 0$
where
\begin{eqnarray}
G_{N+M}^\theta (x_1, \cdots x_{N+M}) &=& T e^{\frac{i}{2}\sum_{I<J}\partial_{x_I}
\wedge \partial_{x_J}} W_{N+M}^0 (x_1, \cdots x_{N+M}) \nonumber \\
:&=&~T~W_{N+M}^\theta (x_1, \cdots x_{N+M})
\label{GNtheta}
\end{eqnarray}
and $W_{N+M}^0$ are the standard Wightman functions for untwisted fields:
\begin{equation}
W_{N+M}^0 (x_1, \cdots x_{N+M}) = \langle 0 | \phi_0 (x_1) \cdots \phi_0(x_{N+M})
|0\rangle.
\end{equation}
It is important that because of translational invariance, the
$W_{N+M}^\theta$ (and hence the $G_{N+M}^\theta$) depend only on coordinate
differences.
 
For simplicity, we have included only matter fields, and that too of one type, in 
(\ref{scatter}). Gauge fields can be included too, but they are not acted on by the twist
exponential in (\ref{GNtheta}).

The scattering matrix element is thus:
\begin{equation}
_\theta \langle -q_N,-q_{N-1},\cdots -q_1; {\rm out}|q_{N+M},q_{N+M-1},\cdots ,q_{N+1};{\rm
  in}\rangle _\theta ~=~\int {\cal{I}} G_{N+M}^\theta(x_1,x_2,\cdots,x_{M+N})
\label{ncLSZ}
\end{equation}
where ${\cal{I}}$ is as defined in (\ref{II}).

On Fourier transforming as in (\ref{LSZformula}), the $\theta_{ij}$
(space-space) part of the twist can be partially integrated. It gives
the usual phase $e^{\frac{i}{2} q_I^i \theta_{ij} q_J^j}$. The time
step-function (in the time-ordering T) in (\ref{GNtheta}) does not
affect this manipulation.

To handle the $\theta_{0i}$ part, consider a typical term
\begin{eqnarray} 
g_{N+M}^\theta (x_1 \cdots x_{N+M}) &=&\theta(x_1^0 - x_2^0) \theta(x_2^0 -
x_3^0) \cdots \theta(x_{N+M-1}^0 - x_{N+M}^0) \nonumber \\
&& e^{\frac{i}{2} \sum_{I<J}
  \partial_{x_I} \wedge \partial_{x_J}} W_{N+M}^0 (x_1, \cdots x_{N+M}).
\label{ncLSZ1}
\end{eqnarray} 
which occurs on expanding the time-ordered product in terms of
retarded products.

The twist here is the product of terms 
\begin{equation} 
e^{\frac{i}{2} [\partial_{x^0_I} \theta \cdot \nabla_J - (\theta \cdot
    \nabla_I) \partial_{x^0_J}]}, \quad I<J, \quad \theta \cdot
    \nabla_J \equiv \theta_{0i} \partial_{x_J^i}
\label{twist}
\end{equation} 
on retaining just $\theta_{0i}$. The coefficient of $\partial_{x^0_I}$
in the exponential is thus
\begin{equation} 
\sum_{J>I} \theta \cdot \nabla_J - \sum_{J<I} \theta \cdot \nabla_J
\end{equation} 
On partial integration in eq(\ref{ncLSZ}) 
$\nabla_J$ becomes $iq_J$ and 
\begin{equation} 
e^{\frac{i}{2} \partial_{x_I^0} (\sum_{J>I} - \sum_{J<I})\theta \cdot
  \nabla_J} \rightarrow e^{-\frac{1}{2} (\sum_{J>I} \theta \cdot q_J -
  \sum_{J<I} \theta\cdot q_J) \partial_{x_I^0}}
\end{equation} 
This translates the $x_I^0$'s according to
\begin{eqnarray} 
x_{I-1}^0 &\rightarrow& x_{I-1}^0 - \frac{1}{2} \left(\sum_{J>I-1}
  \theta \cdot q_J - \sum_{J<I-1} \theta\cdot q_J\right), \\
x_I^0 &\rightarrow& x_I^0 - \frac{1}{2} \left(\sum_{J>I}
  \theta \cdot q_J - \sum_{J<I} \theta\cdot q_J\right), \\
x_{I+1}^0 &\rightarrow& x_{I+1}^0 - \frac{1}{2} \left(\sum_{J>I+1}
  \theta \cdot q_J - \sum_{J<I+1} \theta\cdot q_J\right)
\end{eqnarray} 
Or
\begin{equation} 
x_{I-1}^0 - x_I^0 \rightarrow (x_{I-1}^0 - x_I^0) - \frac{1}{2} \theta
\cdot q_J + \frac{1}{2} \sum_{J \neq I-1,I} \theta \cdot q_J.
\end{equation} 
But $\sum \vec{q}_J = 0$, so that
\begin{equation} 
x_{I-1}^0 - x_I^0 \rightarrow x_{I-1}^0 - x_I^0~ - ~\frac{1}{2} \theta
\cdot q_{I-1}~-~  \frac{1}{2} \theta \cdot q_I
\label{diff1}
\end{equation} 
Similarly,
\begin{equation} 
x_I^0 - x_{I+1}^0 \rightarrow x_I^0 - x_{I+1}^0 ~-~ \frac{1}{2} \theta
\cdot q_I ~-~ \frac{1}{2} \theta \cdot q_{I+1}
\label{diff2}
\end{equation} 
From (\ref{diff1},\ref{diff2}), we see that each $x_I^0$ is (time)
shifted to
\begin{equation} 
x_I^0 + \delta x_I^0, \quad \delta x_I^0 = \delta x_I^0 (q_1, \cdots q_N).
\end{equation} 
where the $\delta x_I^0$ actually depend on the ordering on $x_I^0$.
No further simplification seems possible.

We emphasize the following important observations. 
\begin{itemize}
\item Firstly, (\ref{ncLSZ}) involves only the $\theta^{\mu\nu}=0$
fields in $W_N^0$. So it can be used to map any commutative theory to
noncommutative one, including also the standard model. But special
care is needed to treat gauge fields.  Gauge fields are {\it{not}}
twisted unlike matter fields. As explained elselwhere, this means that
the Yang-Mills tensor is not twisted,
$F^{\mu\nu}_\theta~=~F^{\mu\nu}_0$. But covariant derivatives of
matter fields $\phi_\theta$ are twisted: $(D_\mu\phi)_\theta~=~
(D_\mu\phi)_0~e^{\frac{1}{2}{\overleftarrow{\partial}}\wedge
P}$. where $(D_\mu\phi)_0$ is the untwisted covariant derivative of
the untwisted $\phi_0$.  Thus in correlators $W_N^\theta$, we must use
$(D_\mu\phi)_\theta$ for matter fields, $F^{\mu\nu}_0$ for Yang-Mills
tensor.
\item There are ambiguities in formulating scattering theory. Thus if
we substitute (\ref{phiin}) directly in the LHS of (\ref{LSZformula}),
we get our earlier result \cite{07081379}. This corresponds to putting
the twist factor in (\ref{twist}) outside the symbol T in
(\ref{GNtheta}).  At this moment, lacking a rigorous scattering
theory, we do not know which of these is the correct answer.  In this
connection, we must mention the important work of Buchholz and
Summers\cite{buchholz} which rigorously develops the wedge
localisation ideas of Grosse and Lechner to establish a scattering
theory for two incoming and two outgoing particles. The result
resembles those in our earlier approach\cite{bpq}, but there seem to
be descrepencies in the signs of the momenta in the overall phases.
\end{itemize}

For calculating (\ref{ncLSZ}), we need a formalism for doing
perturbation theory to calculate Wightman functions. Once we have
that, we can calculate the time-ordered product by writing it in terms
of Wightman functions and twist factors.  We show such a calculation
below.
\subsection{Perturbation Theory for Wightman Functions}
Perturbation theory for Wightman functions is available
\cite{ostendorf}. We can also construct this formalism directly. For
free fields, Wightman functions are Gaussian correlated. For example,
\begin{equation}
\langle \phi_F(x_1)\phi_F(x_2)\phi_F(x_3)\phi_F(x_4) \rangle= \langle
\phi_F(x_1)\phi_F(x_2)\rangle \langle\phi_F(x_3)\phi_F(x_4) \rangle
+{\rm permutations}
\label{wightman}
\end{equation}
while for Dirac fields, there are signs attached to the succesive
terms reflecting the signature of the permutations in
$x_1,x_2,x_3,x_4$.  The two-point functions here are well-known. For
example,
\begin{equation}
\langle \phi_F(x_1)\phi_F(x_2)\rangle = \Delta_+(x_1~-~x_2,m^2).
\end{equation}
Now to calculate Wightman functions for interacting fields, we can
expand Heisenberg fields $\phi$ in terms of free fields using
(\ref{heisenberg}) and express the resultant free field correlators in
terms of two point functions.

As an illustration, consider 
\begin{eqnarray}
\lefteqn{\langle T(\phi_F(x_1)\phi_F(x_2)\phi_F(x_3)\phi_F(x_4))
  \rangle =} \nonumber \\
&&\theta(x_1^0-x_2^0)\theta(x_2^0-x_3^0) \theta(x_3^0-x_4^0)
\langle \phi_F(x_1)\phi_F(x_2)\phi_F(x_3)\phi_F(x_4) \rangle + \cdots
\label{wightman1}
\end{eqnarray}
The Wightman function is then given by (\ref{wightman}). We can
similarly calculate the remaining terms in (\ref{wightman1}).

Renormalisation theory for Wightman functions has also been developed
\cite{ostendorf}.
\section{Summary of our Rules for Twisted Quantum Field  Theories}
Our rules for transition from the $\theta^{\mu\nu}=0$ to the
$\theta^{\mu\nu}\neq 0$ theory are simple and definite. Let us focus
on scattering amplitudes. They are given by reduction formulae as in
(\ref{ncLSZ}). They show that to compute scattering amplitudes for
$\theta^{\mu\nu}\neq 0$, we need a formula for twisted Wighman
functions $W_{N+M}^\theta$ in (\ref{GNtheta}) in terms of the untwisted Wightman functions
$W_{N+M}^0$. We have already explained this formula: the passage from
$W_{N+M}^0$ to $W_{N+M}^\theta$ is achieved by twisting all fields except the
gauge fields for unbroken gauge symmetries $H$. Further if
$D(V)^{\theta = 0}$ is the connection with gauge potential $V$ for the
unbroken group $H$, then the covariant derivative for a twisted matter
field $\psi^\theta$ is to be defined as $D_\mu^\theta(V)\psi^\theta :
= (D_\mu(V)^0\psi^0) e^{\frac{1}{2} \overleftarrow{\partial} \wedge
P}$.

This rule preserves the asymptotic conditions and shows that the
spectrum of the theory is unaltered by changing $\theta^{\mu\nu}$. It
covers theories with spontaneous symmetry breakdown as well provided
we have a scheme for treating it for $\theta^{\mu\nu} = 0$.  We remark
in this connection that since twist factors with time derivatives
occur in (\ref{GNtheta}) within the time-ordering symbol and the
amount of time-translation they generate depend on external momenta,
the Gell-Mann-Low formula has to be modified significantly.

\section{Spontaneous Symmetry Breaking}

This topic is of sufficient importance that we discuss it
separately. The consistency of this discussion with what we discussed
earlier will be apparent.

Let us first consider the case of spontaneously broken global
symmetries. Suppose we have a multiplet of quantum fields
$\phi_{i}(x)$ that transforms under the action of (some representation
$D(g)$ of) a symmetry group $G$ according to
\begin{equation} 
\phi_i(x) \rightarrow \phi^g_i(x) = D(g)_{ij} \phi_j(x)
\end{equation} 
If this is a symmetry of the theory, then the quantum charges $Q^a_0$
commute with the full Hamiltonian:
\begin{equation} 
 [Q^a_0, H] =0, \quad a= 1,2, \cdots {\rm dim}\,\, G.
\end{equation} 
These conventionally arise from quantum currents $J^{a,\mu}(x)$ which
are conserved:
\begin{equation} 
\partial_\mu J^{a,\mu}_0 =0
\end{equation} 
where the currents $J^{a,\mu}_0$ are constructed from the quantum fields
$\phi_{i,0}(x)$ and its derivatives.

Given a commutative quantum theory with conserved currents $J^{a,
  \mu}_0$, it is easy to see that in the corresponding noncommutative
  theory (obtained by replacing $\phi_{i,0}(x) \rightarrow
  \phi_{i,\theta}(x) = \phi_{i,0}(x) e^{\frac{1}{2}
  \overleftarrow{\partial} \wedge P}$), the noncommutative
  currents $J^{a,\mu}_\theta (x) = J^{a,\mu}_0(x)e^{\frac{1}{2}
  \overleftarrow{\partial} \wedge P}$ are also conserved:
\begin{equation} 
\partial_\mu J^{a,\mu}_\theta (x) = \partial_\mu J^{a,\mu}_0(x)e^{\frac{1}{2}
  \overleftarrow{\partial} \wedge P} = (\partial_\mu
  J^{a,\mu}_0(x))e^{\frac{1}{2} \overleftarrow{\partial} \wedge P} =
  0.
\end{equation} 
Interestingly, the charges $Q^a_\theta$ in the noncommutative theory
\begin{equation} 
Q^a_\theta = \int d^3 x J^{a,0}_\theta (x) = \int d^3 x J^{a,0}_0 (x)
  e^{\frac{1}{2} \overleftarrow{\partial} \wedge P} = \int d^3 x
  J^{a,0}_0 (x) = Q^a_0
\end{equation} 
are the same as in the commutative case. The last equation follows
using integration by parts for terms involving $\theta_{ij}$ and using
the time independence of the charges of the commutative theory for the
rest.  The charges $Q_0^a, Q_\theta^a$ generate the infinitesimal symmetry
transformations:
\begin{eqnarray} 
[Q^a_0, \phi_{i,0}(x)] &=& \sum_j T^a_{ij}\phi_{j,0}(x), \label{cQ}\\
 {[}Q^a_\theta, \phi_{i,\theta }(x)] &=& \sum_j T^a_{ij} \phi_{j,\theta }(x). \label{ncQ}
\end{eqnarray} 

\subsection{Goldstone's theorem}

Consider the vacuum expectation value of the commutator of the currents
$J^{a,\mu}_\theta(y)$ and the quantum field $\phi_{i,\theta}(x)$:
\begin{equation} 
\langle 0|[J^{a,\mu}_\theta(y), \phi_{i,\theta}(x)]|0\rangle = \langle
0| [e^{\frac{1}{2}\overrightarrow{\partial_y} \wedge P} J^{a,\mu}_0(y),
\phi_0(x)e^{\frac{1}{2} \overleftarrow{\partial_x} \wedge P}] |0\rangle.
\end{equation} 
The first term in the commutator here is 
\begin{equation}
e^{\frac{i}{2}\partial_y\wedge \theta  \partial_x} \langle
0|J_0^{a,\mu}(y)\phi_0(x)|0 \rangle = 
e^{\frac{i}{2}\partial_y\wedge \theta \partial_x} \langle
0|J_0^{a,\mu}(0)\phi_0(x-y)|0 \rangle
\end{equation}
where we have used translational invariance and the fact that
\begin{equation}
\partial_y\wedge \partial_x \langle 0|J_0^\mu(0)\phi_0(x-y)| 0\rangle ~=~-~
\partial_x \wedge \partial_x \langle 0|J_0^\mu(0)\phi_0(x-y) | 0\rangle~=~0
\end{equation}
in the second step. The $\theta^{\mu\nu}$ dependence in the second
term as well disappears in the same manner so that
\begin{equation}
\langle 0|[J^{a,\mu}_\theta(y), \phi_{i,\theta}(x)]|0\rangle = \langle
0|[J^{a,\mu}_0(y), \phi_{i,0}(x)]|0\rangle. 
\end{equation}
This commutator being the same as the one for the corresponding
commutative case, the standard arguments using spectral density and
Lorentz invariance \cite{gsw} can be used to argue for the
existence of massless bosons in the symmetry-broken phase. Following
\cite{weinberg}, we reproduce this argument below.

Summing over intermediate states, and using Lorentz
invariance, the vacuum expectation value of the commutator may be
expressed as
\begin{equation} 
\langle 0|[J^{a,\mu}_0(y), \phi_{i,0}(x)]|0\rangle = \int d^4 p
\left( \rho^{a,\mu}_i(p) e^{-i p \cdot (y-x)} -
\tilde{\rho}^{a,\mu}_i(p) e^{ip \cdot
  (y-x)}\right), 
\end{equation} 
where the spectral densities $\rho^a_i(p), \tilde{\rho}^a_i(p)$ are
defined as
\begin{eqnarray} 
\rho^{a,\mu}_i(p) &=& \sum_N \langle 0| J^{a,\mu}_0(0)|N\rangle \langle
N| \phi_{i,0}(0)|0\rangle \delta^4(p-p_N), \\
\tilde{\rho}^{a,\mu}_i(p) &=& \sum_N \langle 0|\phi_{i,0}(0)|N\rangle \langle
N| J^{a,\mu}_0(0)|0\rangle \delta^4(p-p_N),
\end{eqnarray} 
and $p_N$ is the total four-momentum in the state $\mid N \rangle$.

By Lorentz invariance and non-negativity of energy, these densities are of the form
\begin{eqnarray} 
\rho^{a,\mu}_i(p)&=& p^\mu \rho^a_i (p^2)\theta(p^0), \\
\tilde{\rho}^{a,\mu}_i(p)&=& p^\mu \tilde{\rho}^a_i (p^2)\theta(p^0),
\end{eqnarray} 

which implies that
\begin{eqnarray} 
\lefteqn{\langle 0|[J^{a,\mu}_0(y), \phi_{i,0}(x)]|0\rangle =}
\nonumber \\
&& i \frac{\partial}{\partial y_\mu} \int dM^2 \left( \rho^a_i (M^2)
\Delta_+(y-x;M^2) +\tilde{\rho}^a_i (M^2) \Delta_+(x-y;M^2) \right)
\end{eqnarray} 
where $\Delta_+(x;M^2)$ is the standard two-point Wightman function:
\begin{equation} 
\Delta_+(x;M^2) = \int d\mu(p) e^{-i p \cdot x}, \quad {\rm where}
\quad d\mu(p) = \frac{d^3 p}{2p_0}, \quad p_0 = \sqrt{\vec{p}^2 + M^2}.
\end{equation} 

Since $\Delta_+(x;M^2)$ and $\Delta_+(-x;M^2)$ are equal for $x$
spacelike, we can write, for such $x-y$,
\begin{equation} 
\langle 0|[J^{a,\mu}_0(y), \phi_{i,0}(x)]|0\rangle =
i \frac{\partial}{\partial y_\mu} \int dM^2 \left( \rho^a_i (M^2)+
\tilde{\rho}^a_i (M^2) \right) \Delta_+(y-x;M^2).
\end{equation} 

For spacelike separations, the commutator vanishes, so that 
\begin{equation} 
\rho^a_i (M^2) = -\tilde{\rho}^a_i (M^2)
\end{equation} 
which gives us
\begin{equation} 
\langle 0|[J^{a,\mu}_0(y), \phi_{i,0}(x)]|0\rangle =
i \frac{\partial}{\partial y_\mu} \int dM^2 \rho^a_i(M^2) (\Delta_+(y-x;M^2) -
\Delta_+(x-y;M^2))
\end{equation} 
Now, since the current $J^{a,\mu}_0$ is conserved, we can act by
$\partial/\partial y_\mu$ to get
\begin{equation}
0 = \int dM^2 M^2 \rho^a_i(M^2)(\Delta_+(y-x;M^2) - \Delta_+(x-y;M^2))
\end{equation} 
and thus we get
\begin{equation} 
M^2 \rho^a_i(M^2) = 0. \label{eq1}
\end{equation} 

Now consider the situation when the symmetry is broken. For $\mu =0,
x^0=y^0=t$, 
\begin{equation} 
\langle 0|[J^{a,0}_0(\vec{y},t), \phi_{i,0}(\vec{x},t)]|0\rangle = i
\delta(\vec{y}-\vec{x}) \int dM^2 \rho^a_i(M^2).
\end{equation} 

Integrating and using (\ref{cQ}), we get
\begin{equation} 
\sum_j T^a_{ij}< \phi_{j,0}(x)> = i\int \rho^a_i(M^2). \label{eq2}
\end{equation} 

Eqs. (\ref{eq1}) and (\ref{eq2}) are compatible only if
\begin{equation} 
\rho^a_i(M^2) = i \delta(M^2) \sum_j T^a_{ij}\langle 0|
\phi_{j,0}(0)|0 \rangle 
\end{equation} 
As long as the symmetry is broken, the spectral density $\rho^a_i$ is
proportional to $\delta(M^2)$. Since such a term can arise only in a
theory with massless particles, we are forced to conclude that a
broken symmetry with $T^a_{ij}\langle 0|\phi_{j,\theta}(0)|0\rangle~\neq 0$
requires the existence of a massless particle with the same quantum
numbers as $J^{a,0}_\theta$. These are nothing but the Goldstone
bosons.

\subsection{Spontaneously Broken Local Symmetries \& twisted standard model}
Now given the map between the twisted fields and untwisted ones
(eq. \ref{twistfield}) and our earlier established rules for getting
the correlation functions for the case of $\theta_{\mu\nu}\neq 0$
eqs. {(\ref{GNtheta})} and {(\ref{ncLSZ1})} we can easily see that the Higgs
mechanism will follow with the mass of the gauge boson being identical
to the untwisted case. We can readily understand this result from the fact that
the {\it in} and {\it out} fields completely determine the mass spectrum and they 
remain independent of $\theta_{\mu\nu}$ because of formulae like 
(\ref{phiin}) and (\ref{inout}).

%
%
%
%
The Hamiltonian $P_0~=~H$ and the spatial translation generator for
the twisted standard model are the same as for the case
$\theta^{\mu\nu}~=~0$ What is changed in our LSZ approach are the in
and out fields which are twisted as discussed. Hence scattering
calculations can be based on appropriately modified Wightman functions
as we have already explained.

We will elsewhere calculate specific twisted standard model
cross-sections and examine the new features coming from
non-commutativity.

\section{Final Remarks}

In this paper, we have outlined an approach for calculating the
scattering amplitudes in twisted qft's from untwisted ones using LSZ
formalism. It works in gauge theories with or without spontaneous
breakdown. Implications for the standard model will be presented in a
forthcoming paper.

As remarked earlier, the results for scattering matrix in this
approach differs from the interaction representation perturbation
theory. The reasons for this difference remain to be pinpointed.

In our judgement, since the LSZ approach works with fully interacting
fields and total momentum $P_\mu$ (including also interactions), it is
probably superior to the results based on interaction representation
perturbation theory. It does not change $P_\mu$ in the process of
twisting, but changes just the in- and out- fields appropriately to
account for the twisted statistics.  This change is forced on us when
the coproduct of the Poincar\'e-Hopf algebra is twisted.

In the presence of matter and gauge fields, the coproduct for the
Poincar\'e algebra becomes non-associative and gives rise to a Poincar\'e
- quasi Hopf algebra \cite{BALBABAR}. We will discuss this quasi- Hopf
algebra in detail in another paper.

{\bf Acknowledgments:} The work of APB is supported in part by US-DOE
under grant number DE-FG02-85ER40231 and the Universidad Carlos III
de Madrid. The work of APB and TRG are supported by the DST CP-STIO program.

\end{document}